\makeatletter
\@ifundefined{@parse@version@dash}{%
	\def\@parse@version#1{\@parse@version@0#1}
	\def\@parse@version@#1/#2/#3#4#5\@nil{%
		\@parse@version@dash#1-#2-#3#4\@nil}
	\def\@parse@version@dash#1-#2-#3#4#5\@nil{%
		\if\relax#2\relax\else#1\fi#2#3#4 }
}{}
\makeatother

\makeatletter
\DeclareFontFamily{U}{tipa}{}
\DeclareFontShape{U}{tipa}{m}{n}{<->tipa10}{}
\newcommand{\arc@char}{{\usefont{U}{tipa}{m}{n}\symbol{62}}}%

\newcommand{\arc}[1]{\mathpalette\arc@arc{#1}}

\newcommand{\arc@arc}[2]{%
	\sbox0{$\m@th#1#2$}%
	\vbox{
		\hbox{\resizebox{\wd0}{\height}{\arc@char}}
		\nointerlineskip
		\box0
	}%
}
\makeatother

\documentclass[%
reprint,
superscriptaddress,
amsmath,amssymb,
pre,
showkeys
]{revtex4-2}

\usepackage{color}
\usepackage{graphicx}
\usepackage{dcolumn}
\usepackage{bm}
\usepackage{hyperref}
\newcommand*{\dif}{\mathop{}\!\mathrm{d}}
\newcommand*{\calI}{\mathcal{I}}

\newcommand*{\calR}{\mathcal{R}}

\newcommand*{\calT}{\mathcal{T}}
\newcommand*{\bfb}{{\bf b}}

\newcommand*{\bfe}{{\bf e}}
\newcommand*{\bft}{{\bf t}}
\newcommand*{\bfu}{{\bf u}}
\newcommand*{\bfx}{{\bf x}}
\newcommand*{\bfy}{{\bf y}}
\newcommand*{\bfM}{{\bf M}}
\newcommand*{\bfR}{{\bf R}}
\newcommand*{\bfT}{{\bf T}}

\newtheorem{theorem}{Theorem}
\newtheorem{lemma}{Lemma}

\begin{document}

\title{Theorem for the design of deployable kirigami tessellations with different topologies}

\author{Xiangxin Dang}
\affiliation{State Key Laboratory for Turbulence and Complex Systems, Department of Mechanics and Engineering Science, College of Engineering, Peking University, Beijing 100871, China}
\author{Fan Feng}
\affiliation{Cavendish Laboratory, University of Cambridge, Cambridge CB3 0HE, United Kingdom}
\author{Huiling Duan}
\author{Jianxiang Wang}
\email{jxwang@pku.edu.cn}
\affiliation{State Key Laboratory for Turbulence and Complex Systems, Department of Mechanics and Engineering Science, College of Engineering, Peking University, Beijing 100871, China}
\affiliation{CAPT-HEDPS, and IFSA Collaborative Innovation Center of MoE, College of Engineering, Peking University, Beijing 100871, China}

\date{\today}

\begin{abstract}
The concept of kirigami has been extensively utilized to design deployable structures and reconfigurable metamaterials. Despite heuristic utilization of classical kirigami patterns, the gap between complex kirigami tessellations and systematic design principles still needs to be filled. In this paper, we develop a unified design method for deployable quadrilateral kirigami tessellations perforated on flat sheets with different topologies. This method is based on the parametrization of kirigami patterns formulated as the solution of a linear equation system. The geometric constraints for the deployability of parametrized cutting patterns are given by a unified theorem covering different topologies of the flat sheets. As an application, we employ the design method to achieve desired shapes along the deployment path of kirigami tessellations, while preserving the topological characteristics of the flat sheets. Our approach introduces interesting perspectives for the topological design of kirigami-inspired structures and metamaterials.
\end{abstract}


\maketitle


\section{Introduction}

Kirigami has recently become an emerging paradigm for morphable structures and metamaterials \cite{Zhang2015A, tang2017Programmable, YAN2017Deterministic, celli2018shape, Hu2018Stretchable, Gao2018Two, lipton2018handedness, Yi2018Multistable, jin2020kirigami}.
Programmable deformations can be achieved by prescribing the sizes, orientations, and connections of the cuts perforated on flat sheets.
The bridge between geometric distributions of cuts and the corresponding deployability is the key to understanding kirigami-induced deformation mechanisms.
The design principles of deployable kirigami tessellations have been studied in the literature \cite{rafsanjani2016bistable, Seffen2016k-cones, tang2017Design, zhang2017printing, callens2018flat, yang2018geometry, choi2019programming, Tang2019Programmable, an2020Programmable, choi2021compact}.
The previous works mostly focus on varying the sizes and orientations of cuts under fixed connections on an intact sheet without defects such as holes or cracks, that is, the topology of the sheet is fixed.
The kirigami patterns perforated on flat sheets with flexible topologies are investigated in Refs. \cite{Castle2014Making, Castlee2016Additive, chen2016topological, Jiang2020Freeform, Wang2020Keeping}, where the deployability is realized by folding to close the cutting holes.
Conversely, many other kirigami structures are deployed in the way of opening cuts, among which the quadrilateral tessellations are extensively used to achieve unique properties such as reconfigurability \cite{choi2019programming, rafsanjani2017Buckling}, high stretchability \cite{Cho2014Engineering, rafsanjani2019propagation}, and auxeticity \cite{grima2004negative, shan2015Design}.
It is noted that a theorem for rigidly and flat foldable quadrilateral {\it origami} proved by Tachi \cite{tachi2009generalization} has benefited the inverse design of {\it origami} tessellations \cite{dudte2016programming, Hu2020Constructing}.
However, topological design principles for quadrilateral kirigami tessellations are largely absent.

\begin{figure}[!t]
	\centering
	\includegraphics[width=\columnwidth]{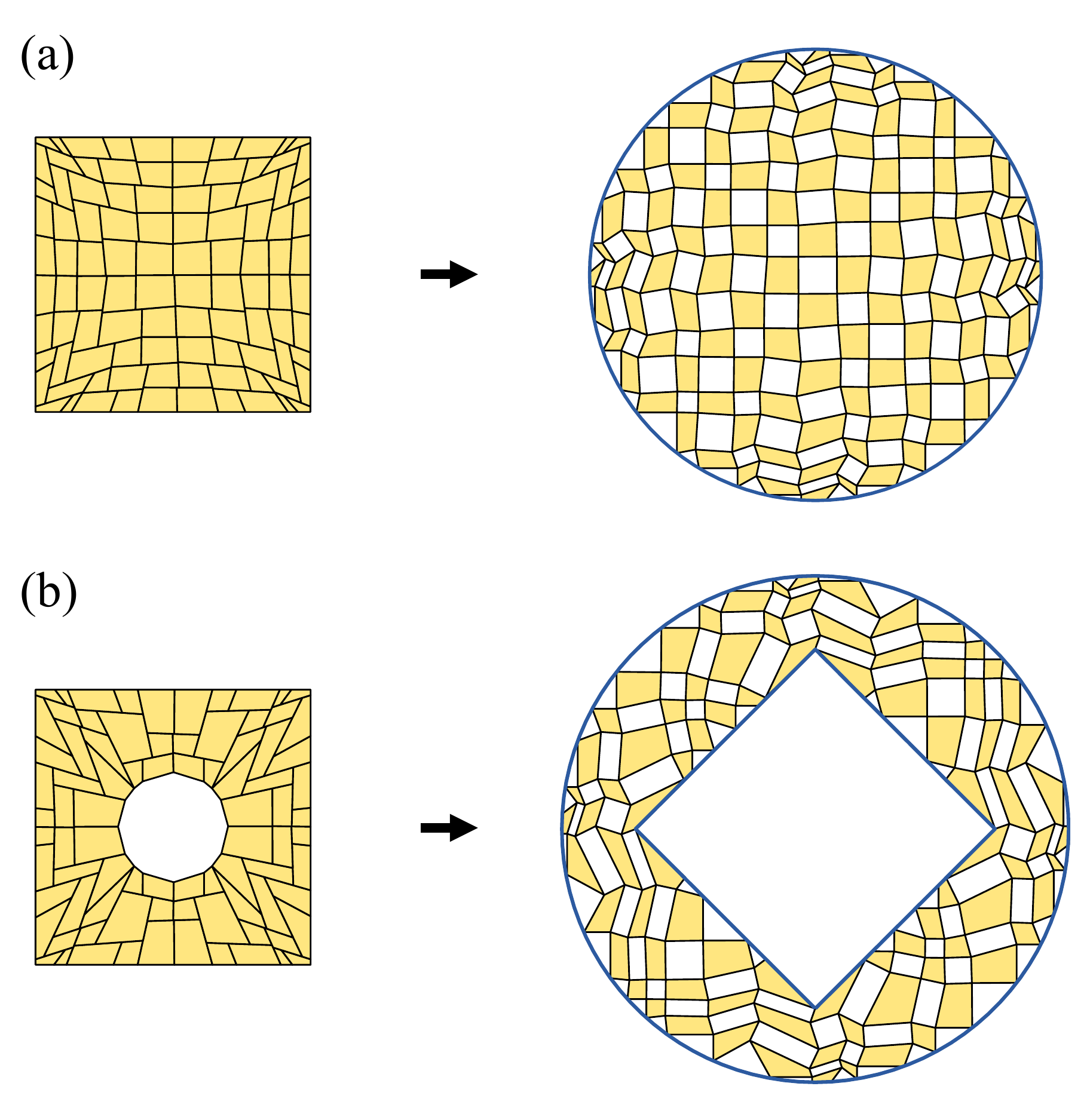}
	\caption{Shape morphing of RDPQK tessellations with different topologies. (a) A genus-0 square RDPQK tessellation that is deployed to approximate a disk. (b) A genus-1 square RDPQK tessellation with a circular hole that is deployed to approximate a disk with a square void. }
	\label{fig:intro}
\end{figure}

\begin{figure*}[!t]
	\centering
	\includegraphics[width=\textwidth]{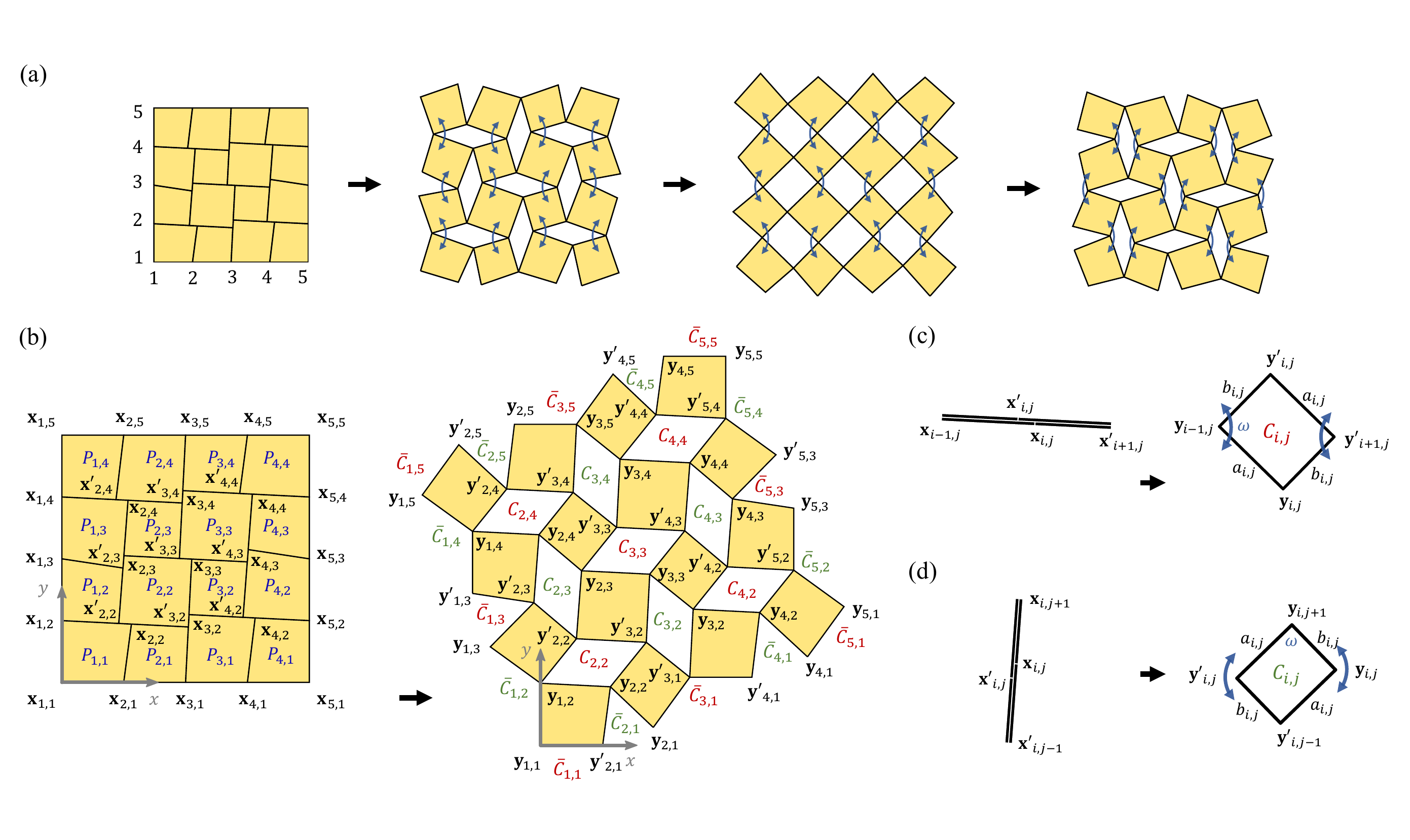}
	\caption{A genus-0 RDPQK tessellation. (a) The deployment of the tessellation with blue arrows representing the expanding directions of opening angles. (b) Geometrical notations of the kirigami pattern (undeformed configuration) and the deployed configurations. The horizontal cuts (red notations, even $i+j$) and vertical cuts (green notations, odd $i+j$) are distributed in a staggered way.  (c) Geometry of the horizontal cuts. (d) Geometry of the vertical cuts.}
	\label{fig:genus0}
\end{figure*}

In this paper, we propose a unified design method for rigidly deployable planar quadrilateral kirigami (RDPQK) tessellations with different topologies. 
Here, the planar quadrilateral kirigami (PQK) tessellations are constructed by cutting flat sheets into arrayed quadrilateral panels connected by flexible hinges at the corners, and \emph{rigid deployability} means that the tessellation can be deployed by rotating the quadrilateral panels around the corners while preserving their sizes, shapes, and connections.
The topology of a sheet to be cut is determined by the value of \emph{genus} \cite{armstrong2013basic}, which equals the number of holes on the sheet.
Specifically, the holes are made on the undeployed state by removing some of the arrayed panels and introducing interior boundaries to the PQK tessellations.
{We will formulate the vertex positions of the quadrilateral kirigami patterns by a set of linear equations, which can be efficiently solved.
Benefiting from the linear formulations, we can optimize the vertex positions to meet the geometric constraints of rigid deployability.
These geometric constraints are given by a deployability theorem, stating that a PQK tessellation is rigidly deployable if there exists one deployed configuration with parallelogram voids of cuts, regardless of the genus.
The theorem greatly simplifies the realization of deployability, because we only need to guarantee the connections of rigid panels at one specific deployed state, and the entire path of rigid deployment will be automatically achieved.
Further, we can inversely design the RDPQK tessellations to fit a predefined deployed shape by incorporating the objectives of shape morphing and the constraints of rigid deployability into an optimization algorithm.
We will also show that the designed RDPQK tessellations are floppy mechanisms with one degree of freedom, so that the deployment can be easily controlled by a mechanical system with actuators.}

This paper is organized as follows.
First, we focus on the parametrization and formulation of genus-0 RDPQK tessellations with no prescribed holes [see Fig.~\ref{fig:intro}(a), left].
Then, the deployability condition for genus-1 RDPQK tessellations [with a hole; see Fig.~\ref{fig:intro}(b), left] is given and proved.
We will demonstrate the effect on the rigid deployability induced from the topology of flat sheets.
Next, the deployability condition is generalized to the case of genus-$n$ and a unified theorem is summarized for the design of RDPQK tessellations with an arbitrary number of holes.
In the end, as an application of the proposed design method, we give an optimization scheme for the shape morphing of RDPQK tessellations.

\section{Genus-$0$ tessellation}
\label{sect:genus0}
The criteria for rigid deployability of a genus-0 PQK tessellation was first studied by Choi {\it et al.} \cite{choi2021compact}.
They proved that a genus-0 PQK tessellation is rigidly deployable when the voids of cuts are rhombuses.
This criteria has been generalized in our recent work \cite{dang2021a} with the following necessary and sufficient condition:

\begin{lemma}[Genus-$0$ deployability]
	A genus-0 PQK tessellation is rigidly deployable if and only if all the cuts form parallelogram voids.
	\label{lemma1}
\end{lemma}
Lemma~\ref{lemma1} shows that the rigid deployability of a genus-0 PQK tessellation can be achieved by constraining the shapes of opening cuts to be parallelograms.
This lemma was stated as a corollary of a compatibility theorem of spherical kirigami tessellations, and proved as the degeneracy of spherical geometry in Ref. \cite{dang2021a}.
Here we provide a straightforward proof in Appendix \ref{ap:lemma1} directly from the viewpoint of planar geometry.

As a starting point of the unified design method, we parametrize the cutting patterns of genus-0 RDPQK tessellations and formulate the rigid motions of the quadrilateral panels.
The genus-$n$ RDPQK tessellations can be formulated similarly.
Before the formulations, we introduce some notations to characterize the configurations of RDPQK tessellations.
Figure~\ref{fig:genus0}(a) demonstrates the deployment of a $4\times4$ square RDPQK tessellation.
The flat sheet is divided into arrayed quadrilateral panels by interwoven cuts---each cut is intersected by its two neighbors and is split into four segments with two vertices on the cut.
The notations of geometrical elements (panels, cuts, and vertices) for the kirigami pattern and deployed configuration are illustrated in Fig.~\ref{fig:genus0}(b).
Collectively, we use $P_{i,j}$ to denote the panel of the $i$th column and the $j$th row, and the cut between panels $P_{i-1,j-1}$, $P_{i,j-1}$, $P_{i-1,j}$, and $P_{i,j}$ is denoted by $C_{i,j}$.
In the undeployed kirigami configuration, the vertices on the cut $C_{i,j}$ are denoted by $\bfx_{i,j}$ and $\bfx'_{i,j}$, which move to $\bfy_{i,j}$ and $\bfy'_{i,j}$ upon deployment, respectively.
We can classify the cuts into two types by the directions in which they link the vertices, i.e., the cut linking $\bfx_{i-1,j}$ and $\bfx'_{i+1,j}$ is referred to as a horizontal cut, and the one linking $\bfx'_{i,j-1}$ and $\bfx_{i,j+1}$ is a vertical cut, as shown in Figs.~\ref{fig:genus0}(c) and \ref{fig:genus0}(d).
Besides, we add extra boundary cuts $\bar C_{i,j}$ around the outline of the tessellation, which are also classified according to the horizontal or vertical orientations, as shown in Fig.~\ref{fig:genus0}(b).
Altogether, the horizontal and vertical cuts (including the boundary ones) are arranged as a staggered array on the tessellation.
{The array-like feature inspires us to define a {\it {matrix of topology}} with components $\pm 1$ and $\pm 2$ to uniquely represent the staggered arrangement of cuts, i.e.,}
\begin{equation}
	\bfT^0_{4\times4}=\left[
	\begin{array}{rrrrr}
		-1 &     -2 &     -1 &     -2 &     -1 \\
		-2 &      1 &      2 &      1 &     -2 \\
		-1 &      2 &      1 &      2 &     -1 \\
		-2 &      1 &      2 &      1 &     -2 \\
		-1 &     -2 &     -1 &     -2 &     -1
	\end{array}
	\right].
\end{equation}
Here the superscript 0 represents the genus of this tessellation, and the subscript $4\times4$ is the number of panels.
The components $\pm1$ indicate that vertices $\bfx_{i,j}$, $\bfx'_{i,j}$ belong to horizontal cuts, and $\pm2$ correspond to vertical cuts.
The positive and negative signs represent interior and boundary cuts, respectively.
In general, for a genus-0 RDPQK tessellation with $M\times N$ panels, the {matrix of topology} $\bfT^0_{M\times N}$ is an $(M+1)\times (N+1)$ matrix defined by
\begin{equation}
	t_{i,j} = \left\{
	\begin{aligned}
		& \sigma_{i,j}1 & & {\rm for~~even}~~i+j\\
		& \sigma_{i,j}2 & & {\rm for~~odd}~~i+j
	\end{aligned}
	\right.,
	\label{eq:tij}
\end{equation}
for $i=1,2,...,M+1$ and $j=1,2,...,N+1$, in which the sign function $\sigma_{i,j}=+$ for $i=2,...,M$ and $j=2,...,N$, and $\sigma_{i,j}=-$ for $i=1,M+1$ or $j=1,N+1$.
Note that the tessellations discussed here are restricted to having at least three columns and rows of panels (i.e., $M,N\geq3$).
The PQK tessellations with two columns or rows are actually always rigidly deployable even if the voids of cuts are arbitrary quadrilaterals.
We exclude these trivial cases for brevity in this paper.

\begin{figure*}[!t]
	\centering
	\includegraphics[width=\textwidth]{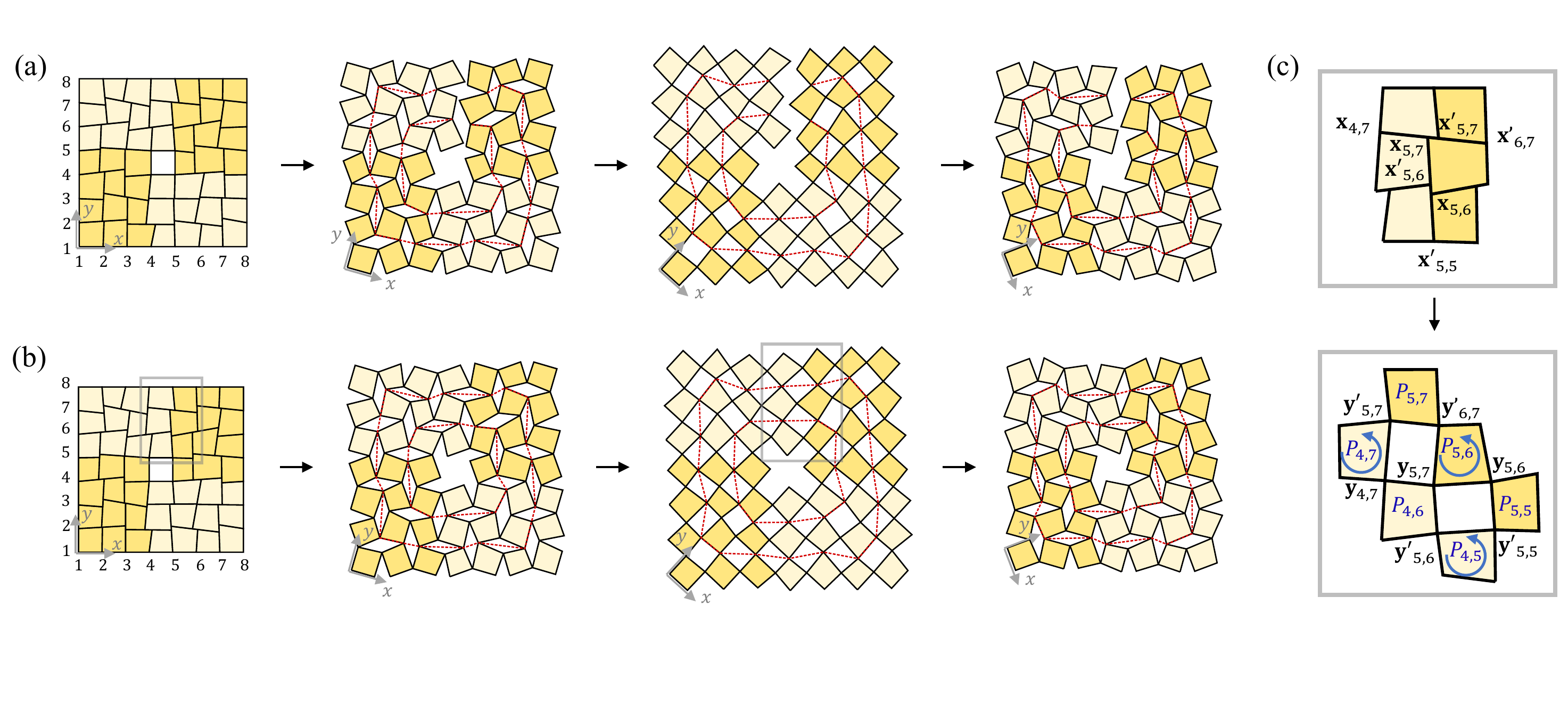}
	\caption{The genus-1 PQK tessellations. Each tessellation can be regarded to be composed of four genus-0 RDPQK tessellations (distinguished by colors of light and dark yellow).  (a) A genus-1 PQK tessellation that is not rigidly deployable. The voids of cuts surrounding a hole cannot form closed rings of parallelograms. (b) A genus-1 RDPQK tessellation. The voids of cuts form two parallelogram rings highlighted by red dotted lines. (c) Notations of the panels above the hole extracted by the boxes from the genus-1 RDPQK tessellation. The blue arrows on the panels represent directions of rotation.}
	\label{fig:genus1}
\end{figure*}

According to Lemma~\ref{lemma1}, the voids of cuts form parallelograms upon deployment of genus-0 RDPQK tessellations.
We denote the side lengths of the parallelogram voids by $a_{i,j}$ and $b_{i,j}$ [see Figs.~\ref{fig:genus0}(c) and \ref{fig:genus0}(d)]; then the aspect ratio for each cut $C_{i,j}$ is defined by $r_{i,j}=b_{i,j}/a_{i,j}$.
If we fix the boundary vertices as $\bfx_{i,j}^{\rm bound}$ with $t_{i,j}<0$ for the kirigami patterns, the positions of interior vertices $\bfx_{i,j}$ and $\bfx'_{i,j}$ with $t_{i,j}>0$ can be uniquely solved by the following linear equation system (LES),
\begin{equation}
	\begin{aligned}
		& (1-r_{i,j})\bfx'_{i+1,j}-\bfx_{i,j}+r_{i,j}\bfx_{i-1,j}=0, & t_{i,j}=1,\\
		& (1-r_{i,j})\bfx_{i-1,j}-\bfx'_{i,j}+r_{i,j}\bfx'_{i+1,j}=0, & t_{i,j}=1,\\
		& (1-r_{i,j})\bfx_{i,j+1}-\bfx_{i,j}+r_{i,j}\bfx'_{i,j-1}=0, & t_{i,j}=2,\\
		& (1-r_{i,j})\bfx'_{i,j-1}-\bfx'_{i,j}+r_{i,j}\bfx_{i,j+1}=0, & t_{i,j}=2,
	\end{aligned}
	\label{eq:pl-linEqSys}
\end{equation}
along with the boundary conditions
\begin{equation}
	\bfx_{i,j}=\bfx'_{i,j}=\bfx_{i,j}^{\rm bound}, ~~~~ t_{i,j}<0.
	\label{eq:pl-linEqSysBC}
\end{equation}
The equations for $t_{i,j}=1$ and $2$ in Eq.~(\ref{eq:pl-linEqSys}) represent the collinear constraints of vertices on horizontal and vertical cuts, respectively.
Equations (\ref{eq:pl-linEqSys}) and (\ref{eq:pl-linEqSysBC}) indicate that we can parametrize the cutting pattern of a genus-0 RDPQK tessellation by the aspect ratios $r_{i,j}$ for cuts $C_{i,j}$ and positions of boundary vertices $\bfx_{i,j}^{\rm bound}$.
For example, the tessellation in Fig.~\ref{fig:genus0} is constructed by setting $r_{i,j}=0.45$ and uniformly distributing the boundary vertices on a square sheet.
Remarkably, the kirigami pattern can be determined efficiently by solving the LES with standard numerical methods, which benefits the inverse design such as the optimization of shape morphing in Sec.~\ref{sect:shape}.

The parallelogram voids of cuts also lead to a concise formulation of deployment  for genus-0 RDPQK tessellations.
As shown in Fig.~\ref{fig:genus0}(b), first, the motion of a panel $P_{i,j}$ with $|t_{i,j}|=1$ is a pure translation relative to $P_{1,1}$.
Second, it can be observed that the opening angles at a common vertex of adjacent parallelogram voids are complementary angles, and therefore, each panel $P_{i,j}$ with $|t_{i,j}|=2$ rotates by the same angle.
In a word, the deployment has one degree of freedom that can be parametrized by a reference opening angle $\omega$ as illustrated in Figs.~\ref{fig:genus0} (c) and \ref{fig:genus0}(d).
The rotation matrix on a plane is given by
\begin{equation}
	\bfR^\omega=
	\left[
	\begin{array}{rr}
		\cos \omega & -\sin \omega \\
		\sin \omega & \cos \omega
	\end{array}
	\right].
\end{equation}
We can derive the rigid transformations $\calR_{i,j}^\omega$ of the motion (combinations of rotations and translations) for each panel $P_{i,j}$ iteratively as follows:
\begin{itemize}
	\item[1)]
	Fix the first panel $P_{1,1}$ by the identity transformation $\calI$:
	\begin{equation}
		\calR_{1,1}^\omega=\calI.
		\label{eq:pl-trans-1}
	\end{equation}
	\item[2)]
	If a panel $P_{i,j}$ has no neighbor below it, calculate the motion relative to its left panel:
	\begin{equation}
		\begin{aligned}
			\calR_{i,j}^\omega\bfx &= (\bfx-\bfx'_{i,j})+\calR_{i-1,j}^\omega\bfx'_{i,j}, & |t_{i,j}|=1, \\
			\calR_{i,j}^\omega\bfx &= \bfR^\omega(\bfx-\bfx_{i,j+1})+\calR_{i-1,j}^\omega\bfx_{i,j+1}, & |t_{i,j}|=2,
		\end{aligned}
		\label{eq:pl-trans-2}
	\end{equation}
	for $t_{i,j}<0$ and $t_{i+1,j}<0$.
	\item[3)]
	If a panel $P_{i,j}$ has a neighbor below it, calculate the motion relative to the panel below:
	\begin{equation}
		\begin{aligned}
			\calR_{i,j}^\omega\bfx &= (\bfx-\bfx'_{i+1,j})+\calR_{i,j-1}^\omega\bfx'_{i+1,j}, & |t_{i,j}|=1, \\
			\calR_{i,j}^\omega\bfx &= \bfR^\omega(\bfx-\bfx_{i,j})+\calR_{i,j-1}^\omega\bfx_{i,j}, & |t_{i,j}|=2, \\
		\end{aligned}
		\label{eq:pl-trans-3}
	\end{equation}
	for $t_{i,j}>0$ or $t_{i+1,j}>0$.
\end{itemize}
Consequently, the displacement of any vertex $\bfx$ on the panel $P_{i,j}$ can be calculated by
\begin{equation}
	\bfy=\calR_{i,j}^\omega\bfx,~\bfx\in P_{i,j}.
	\label{eq:pl-deformation}
\end{equation}
For a genus-0 RDPQK tessellation, the links of corresponding corners of adjacent panels are preserved upon deployment.
In other words, the displacement of a vertex is the same no matter which panel the calculation is based on, e.g., $\bfy'_{3,3}=\calR_{2,3}^\omega\bfx'_{3,3}=\calR_{3,3}^\omega\bfx'_{3,3}$.
This fact follows the rigid deployability depicted in Lemma~\ref{lemma1}.
However, we will show that it is not always the case for genus-$n$ PQK tessellations ($n\geq1$) with parallelogram voids of cuts.

\begin{figure*}[!t]
	\centering
	\includegraphics[width=\textwidth]{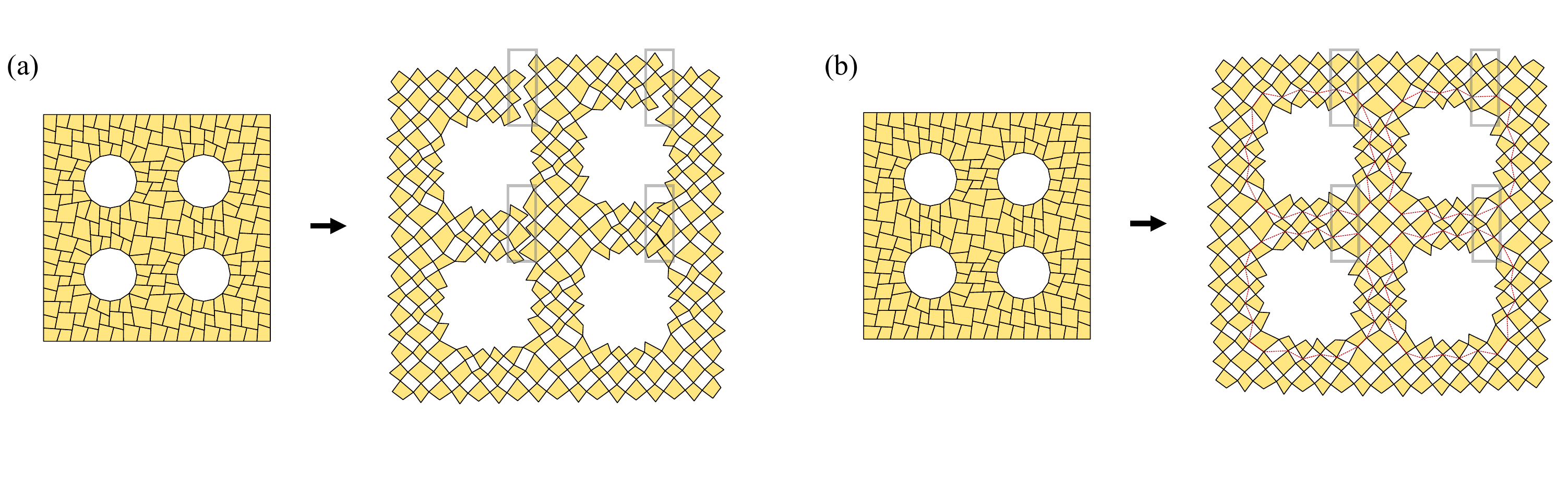}
	\caption{The genus-4 PQK tessellations with circular interior boundaries. (a) A genus-4 PQK tessellation with aspect ratios $r_{i,j}=0.4$. The gray boxes frame four areas of split voids of cuts, which  means this tessellation is not rigidly deployable. (b) A genus-4 RDPQK tessellation optimized from $\bar r_{i,j}=0.4$. The voids of cuts surrounding each hole can form a closed ring of parallelograms, highlighted by the red dotted lines.}
	\label{fig:genus4}
\end{figure*}

\section{Genus-$1$ tessellation}
\label{sect:genus1}

Moving beyond the genus-0 tessellations, we add one or more interior boundaries to modify the topology of flat sheets (i.e., making holes).
This modification can be expressed clearly with the {matrices of topology}.
For example, consider the $7\times7$ PQK pattern with a hole replacing the panel at the center, as illustrated in Fig.~\ref{fig:genus1} (a).
We reverse the signs of the four components at the center of the {matrix of topology} $\bfT^0_{7\times7}$ defined by Eq.~(\ref{eq:tij}), and obtain the genus-1 {matrix of topology}
\begin{equation}
	\bfT^1_{7\times7}=
	\left[
	\begin{array}{rrrrrrrr}
		-2 &     -1 &     -2 &     -1 &     -2 &     -1 &     -2 &     -1 \\
		-1 &      2 &      1 &      2 &      1 &      2 &      1 &     -2 \\
		-2 &      1 &      2 &      1 &      2 &      1 &      2 &     -1 \\
		-1 &      2 &      1 &     -2 &     -1 &      2 &      1 &     -2 \\
		-2 &      1 &      2 &     -1 &     -2 &      1 &      2 &     -1 \\
		-1 &      2 &      1 &      2 &      1 &      2 &      1 &     -2 \\
		-2 &      1 &      2 &      1 &      2 &      1 &      2 &     -1 \\
		-1 &     -2 &     -1 &     -2 &     -1 &     -2 &     -1 &     -2 \\
	\end{array}
	\right].
	\label{eq:tm7}
\end{equation}
The negative components $t_{4,4}$, $t_{4,5}$, $t_{5,4}$, and $t_{5,5}$ indicate that the associated vertices $\bfx_{i,j}$ and $\bfx'_{i,j}$ for $i,j=4,5$ are converted to boundary vertices and the panel $P_{4,4}$ is removed.
To keep the configurations unified with genus-0 tessellations discussed in Sec.~\ref{sect:genus0}, the distance between the exterior and interior boundaries has been limited to be not fewer than three panels (the same below).
As a result, it is necessary that the voids of cuts are parallelograms to guarantee the rigid deployability of genus-1 PQK tessellations.
Therefore, the genus-1 patterns can also be determined by solving Eqs.~(\ref{eq:pl-linEqSys}) and (\ref{eq:pl-linEqSysBC}) if the aspect ratios $r_{i,j}$ of cuts are given and the positions of vertices on both the exterior and interior boundaries are fixed.
Specifically, we apply randomized $r_{i,j}\in[0.4,0.6]$ to construct the PQK tessellation in Fig.~\ref{fig:genus1}(a).
Besides, the deployment process of this tessellation is obtained according to the same rules in Eqs.~(\ref{eq:pl-trans-1})--(\ref{eq:pl-deformation}).
However, what we observe is that the calculated deformations are not continuous as adjacent panels above the hole are disconnected, even though the voids of cuts are all parallelograms.
This phenomenon shows that the tessellation in Fig.~\ref{fig:genus1}(a) is not rigidly deployable and Lemma~\ref{lemma1} is not applicable to genus-$n$ pattern when $n\geq1$.

Now we turn to the constraints of the rigid deployability additionally induced from the modification of topology.
A genus-1 RDPQK tessellation is illustrated in Fig.~\ref{fig:genus1}(b).
Rigid deployability requires that the connections of panels remain unchanged upon deployment.
In other words, all the voids of cuts should be intact parallelograms along the deploying path.
Before elaborating how we obtain the kirigami pattern, we present a  Lemma on the deployability of genus-1 tessellations which will greatly simplify the formulations:
\begin{lemma}[Genus-$1$ deployability]
	A genus-$1$ PQK tessellation is rigidly deployable if and only if there exists a deployed state with all the cuts forming parallelogram voids.
	\label{lemma2}
\end{lemma}
The proof of Lemma~\ref{lemma2} is based on the observation that the voids of cuts can form rings of parallelograms surrounding the interior boundary, and these rings remain closed upon deployment as shown in Fig.~\ref{fig:genus1}(b).
Details of the proof are provided in Appendix \ref{ap:lemma2}.
The key point of Lemma~\ref{lemma2} is that we only need to ensure the connectivity of the panels at one specific deployed configuration (say $\omega=\tilde\omega$), and then the panels will always remain connected throughout the deploying process.
Specifically, for tessellations with a {matrix of topology} $\bfT^1_{7\times7}$ in Eq.~(\ref{eq:tm7}), the connectivity can be expressed by $\calR_{4,5}^{{\tilde\omega}}\bfx'_{5,5} = \calR_{5,5}^{{\tilde\omega}}\bfx'_{5,5}$, $\calR_{4,6}^{{\tilde\omega}}\bfx_{5,7} = \calR_{5,6}^{{\tilde\omega}}\bfx_{5,7}$, and $\calR_{4,7}^{{\tilde\omega}}\bfx'_{5,7} = \calR_{5,7}^{{\tilde\omega}}\bfx'_{5,7}$, as shown in Fig.~\ref{fig:genus1}(c).
Keeping in mind that the motions of panels are obtained from Eqs.~(\ref{eq:pl-trans-1})--(\ref{eq:pl-deformation}),
we have the following relationships: $\calR_{4,7}^{{\tilde\omega}}\bfx'_{5,7} - \calR_{5,7}^{{\tilde\omega}}\bfx'_{5,7} = \calR_{4,6}^{{\tilde\omega}}\bfx_{5,7} - \calR_{5,6}^{{\tilde\omega}}\bfx_{5,7} = \calR_{4,5}^{{\tilde\omega}}\bfx'_{5,5} - \calR_{5,5}^{{\tilde\omega}}\bfx'_{5,5}$.
As a result, these expressions for connections are equivalent to each other.
That is to say, the rigid deployability of a genus-1 RDPQK tessellation essentially introduces two additional constraints to the kirigami patterns.
These two constraints are consistent with Eqs.~(\ref{eq:ring-3}) and (\ref{eq:ring-3-2}) derived in the proof of Lemma~\ref{lemma2}.

Since the connections of panels discussed above are nonlinear constraints with respect to the aspect ratios of cuts, we use an optimization method to find RDPQK patterns.
In general, if we remove the interior panels from columns $I_1$ to $I_2$ and rows $J_1$ to $J_2$ of an $M\times N$ PQK tessellation, the interior boundary can be indexed by $t_{i,j}<0$ for $i=I_1,I_1+1,...,I_2$, $j=J_1,J_2$, and $i=I_1,I_2$, $j=J_1,J_1+1,...,J_2$ in the {matrix of topology} $\bfT^1_{M\times N}$ [e.g., $I_1, J_1=4$ and $I_2, J_2=5$ in Eq.~(\ref{eq:tm7})].
Then the distance between vertices to be connected is given as follows:
\begin{equation}
	\begin{aligned}
		& d_{I_2,J_2}(r_{i,j},\bfx^{\rm bound}_{i,j},\tilde\omega)\\
		= & \left\{
		\begin{aligned}
			& \|\calR^{\tilde\omega}_{I_2-1,J_2}\bfx'_{I_2,J_2} - \calR^{\tilde\omega}_{I_2,J_2}\bfx'_{I_2,J_2}\|, & t_{I_2,J_2}=-1\\
			& \|\calR^{\tilde\omega}_{I_2-1,J_2}\bfx_{I_2,J_2+1} - \calR^{\tilde\omega}_{I_2,J_2}\bfx_{I_2,J_2+1}\|, & t_{I_2,J_2}=-2\\
		\end{aligned}
		\right..
	\end{aligned}
	\label{eq:connect-1}
\end{equation}
We determine a genus-1 RDPQK tessellation by solving the optimization problem:
\begin{equation}
	\begin{aligned}
	& \min_{r_{i,j}} \sum_{i,j}{(r_{i,j}-\bar r_{i,j})^2}\\
	& \text{ \rm subject to } [d_{I_2,J_2}(r_{i,j},\bfx^{\rm bound}_{i,j},\tilde\omega)]^2=0,
	\end{aligned}
	\label{eq:opt-1}
\end{equation}
where the aspect ratios $r_{i,j}$ are optimized to be close to the initial values $\bar r_{i,j}$.
For example, the tessellation in Fig.~\ref{fig:genus1}(b) is obtained from the optimization initialized by the aspect ratios in Fig.~\ref{fig:genus1}(a).
In the optimization, we control the connectivity of panels for the deployed configuration at $\tilde\omega=0.5\pi$ (the same below).
Then the panels above the hole can also be connected at other states of deployment, which verifies the lemma of genus-1 deployability.
Besides, it can be observed that the kirigami patterns of these two tessellations are quite similar.
However, the optimized tessellation is rigid deployable, while the initial one cannot preserve the connectivity of panels.
The obvious difference in the deformed configurations for almost the same kirigami patterns is induced from the high nonlinearity of the finite rotations of panels.

\section{Genus-$n$ tessellation}

Having formulated the genus-1 RDPQK tessellations, we go further to treat the general cases of genus-$n$ ($n\geq 1$).
The {matrix of topology} $\bfT^K_{M\times N}$ with $K$ disjoint interior boundaries (no closer than three panels to each other) can be constructed based on $\bfT^0_{M\times N}$ by reversing the sign of $t_{i,j}$ for $i=I_1^k,I_1^k+1,...,I_2^k$, $j=J_1^k,J_1^k+1,...,J_2^k$, and $k=1,2,...,K$, then setting $t_{i,j}=0$ for $i=I_1^k+1,I_1^k+2,...,I_2^k-1$, $j=J_1^k+1,J_1^k+2,...,J_2^k-1$ (if $I_2^k-I_1^k\geq 2$ and $J_2^k-J_1^k\geq 2$), and $k=1,2,...,K$, where $(I_1^k,J_1^k)$ and $(I_2^k,J_2^k)$ are the lowest and highest indexes of panels removed for the $k$th hole, respectively.
In this way, the minus components represent boundary vertices, and the zero components stand for the vertices inside interior boundaries, which need to be removed from the tessellations.
For example, Fig.~\ref{fig:genus4}(a) illustrates a genus-4 PQK tessellation with exterior boundary vertices uniformly distributed on a square sheet, and interior boundary vertices on four arrayed circles.
We obtain the kirigami pattern by assigning aspect ratios $r_{i,j}=0.4$ and solving Eqs.~(\ref{eq:pl-linEqSys}) and (\ref{eq:pl-linEqSysBC}) based on the {matrix of topology} $\bfT^4_{17\times17}$ in Eq.~(\ref{eq:tm17}).
The deployed configuration [see Fig.~\ref{fig:genus4}(a) right] determined by Eqs.~(\ref{eq:pl-trans-1})--(\ref{eq:pl-deformation}) has four bands of split voids of cuts, indicating that this tessellation is not rigidly deployable.

The condition for rigid deployability of genus-$n$ PQK tessellations can be given by the following theorem :
\begin{theorem}[Genus-$n$ deployability]
	A genus-$n$ ($n\geq0$) PQK tessellation is rigidly deployable if and only if there exists a deployed state with all the cuts forming parallelogram voids.
	\label{theorem1}
\end{theorem}
Theorem~\ref{theorem1} is equivalent to Lemmas~\ref{lemma1} and \ref{lemma2} for $n=0$ and $1$, respectively.
Besides, for $n\geq2$, this theorem can be directly verified by applying the proof of Lemma~\ref{lemma2} to each of the holes.
Similar to the genus-1 case, we optimize the aspect ratios $r_{i,j}$ to construct closed rings of parallelograms around each hole while minimizing the variations from initial aspect ratios $\bar r_{i,j}$:
\begin{equation}
	\begin{aligned}
	& \min_{r_{i,j}} \sum_{i,j}{(r_{i,j}-\bar r_{i,j})^2} \\
	& \text{ \rm subject to } [d_{I_2^k,J_2^k}(r_{i,j},\bfx^{\rm bound}_{i,j},\tilde\omega)]^2=0.
	\end{aligned}
	\label{eq:opt-2}
\end{equation}
For example, the RDPQK tessellation in Fig.~\ref{fig:genus4}(b) is obtained from initial aspect ratios $\bar r_{i,j}=0.4$.
Note that although we only demonstrate an example with arrayed circular holes here, the validity of Theorem~\ref{theorem1} does not rely on the shapes of the holes and flat sheets, because the proof of Lemma~\ref{lemma2} in Appendix \ref{ap:lemma2} is based on a general ring of parallelograms independent of the interior or exterior boundaries.
Therefore, the theorem describes the rigid deployability of quadrilateral kirigami in the sense of different topologies.

\section{Inverse design}
\label{sect:shape}
Theorem~\ref{theorem1} stipulates the geometrical constraints that an RDPQK tessellation should satisfy.
Under these constraints, the aspect ratios and locations of boundary vertices can further be optimized to inversely design the kirigami patterns and achieve desired shapes along the deploying path.
To this end, we minimize the distance between outermost vertices on a deployed tessellation and a given target curve.
Specifically, we solve the following optimization problem
\begin{equation}
	\begin{aligned}
		& \min_{r_{i,j},~\bfx^{\rm bound}_{i,j},~\calT} [h(\bfy^{\rm control}_s;\calT)]^2 \\
		& \text{ \rm subject to } [d_{I_2^k,J_2^k}(r_{i,j},\bfx^{\rm bound}_{i,j},\tilde\omega)]^2=0,
	\end{aligned}
	\label{eq:opt-3}
\end{equation}
where $h$ is the function of a target curve, and $\bfy^{\rm control}_s$ are deployed vertices that we aim to control to match the target curve, in which the indices $s=1,2,...,S$, and $S$ is the total number of the control vertices.
Besides, we apply an affine transformation $\calT$ defined by $\calT(\lambda,\alpha,\bfb)\bfx=\lambda\bfR^\alpha\bfx+\bfb$, which is used to adjust the size and orientation of the target curves.

Figure~\ref{fig:intro}(a) illustrates a square genus-0 tessellation that approximates a circle $h^{\rm circle}(\bfy;\calT) = \|\calT\bfy\|-1$.
We prescribe the control vertices $\bfy^{\rm control}_s$ on the target curve as the outermost ones $\bfy_{i,j}$ and $\bfy'_{i,j}$ for $t_{i,j}=-1$ with $i=1,11$, and $t_{i,j}=-2$ with $j=1,11$.
The optimization is initialized by regular aspect ratios $\bar r_{i,j}=0.5$ and uniformly distributed boundary vertices on a square.
In order to preserve the outline shape of the sheet, we only allow the vertices $\bfx^{\rm bound}$ to slide on the square boundary.
The optimized tessellation can be rigidly deployed to approximate the target circle precisely at $\tilde\omega=0.5\pi$.
In addition, the optimization framework can be harnessed to design general genus-$n$ RDPQK tessellations as well.
For example, the square genus-1 tessellation in Fig.~\ref{fig:intro}(b) is deployed to fit a circle $h^{\rm circle}(\bfy;\calT_1) = \|\calT_1\bfy\|-1$, while the interior boundary approximates a square $h^{\rm square}(\bfy;\calT_2)$.
The function of the square curve is given by $h^{\rm square}(\bfy;\calT_2) = (|\bfe_1\cdot\calT_2\bfy|-1)(|\bfe_2\cdot\calT_2\bfy|-1)+\varphi(|\bfe_1\cdot\calT_2\bfy|-1)+\varphi(|\bfe_2\cdot\calT_2\bfy|-1)$, in which $\bfe_1=(1,0,0)$, $\bfe_2=(0,1,0)$, and $\varphi$ is the ramp function defined by $\varphi(x)=x$ for $x>0$ and $\varphi(x)=0$ for $x\leq0$.
This tessellation is built on the {matrix of topology} $\bfT^1_{10\times10}$ in Eq.~(\ref{eq:tm10}).
We prescribe the same control vertices (denoted by $\bfy^{\rm ext}_{s_1}$) on the exterior boundary as the genus-0 case, and select the interior control vertices $\bfy^{\rm int}_{s_2}$ in an analogous manner.
The optimization is performed by simultaneously minimizing $h^{\rm circle}(\bfy^{\rm ext}_{s_1};\calT_1)$ and $h^{\rm circle}(\bfy^{\rm int}_{s_2};\calT_2)$ in the framework of Eq.~(\ref{eq:opt-3}).
The obtained tessellation is rigidly deployable since the connectivity of panels is preserved by $d_{I_2^k,J_2^k}^2=0$ in the optimization.

\section{Conclusions}
In conclusion, we demonstrate a unified design strategy for RDPQK tessellations perforated on flat sheets with an arbitrary number of holes.
{The effectiveness of the design strategy is attributed to two main factors. First, the kirigami patterns are formulated by an LES that can be efficiently solved via standard numerical methods. Second, the deployability theorem reduces the problem of achieving rigidly deployable deformations from the entire deploying path to a specific deployed configuration.}
Remarkably, this theorem reveals the deployability property of RDPQK tessellations that is independent of the topologies of the flat sheets to be cut.
In addition to shape morphing, the proposed optimization framework can also be used to design kirigami tessellations with other properties such as tunable porosity and programmable Poisson's ratio.
{As future work, we envision a generalization of the current theorem to three dimensional kirigami. 
For instance, Ara\'ujo {\it et al.} \cite{Araujo2018Finding} used network topology to design polyhedral kirigami, which was later extended by Melo {\it et al.} \cite{melo2020optimal} to study the deployable dynamics. But a deployability theorem incorporating the genus number of the  kirigami in three dimension is still absent, which is a promising subject to be investigated.}
Finally, we hope that our study will benefit the applications of kirigami-inspired structures and metamaterials based on various topologies.

\begin{acknowledgments}
	X.D., H.D., and J.W. thank the National Natural Science Foundation of China (Grants No.~11991033, No.~91848201, and No.~11521202) for support of this work.
\end{acknowledgments}

\appendix

\section{Proof of  Lemma~\ref{lemma1}}
\label{ap:lemma1}

\begin{figure}[!t]
	\centering
	\includegraphics[width=0.5\textwidth]{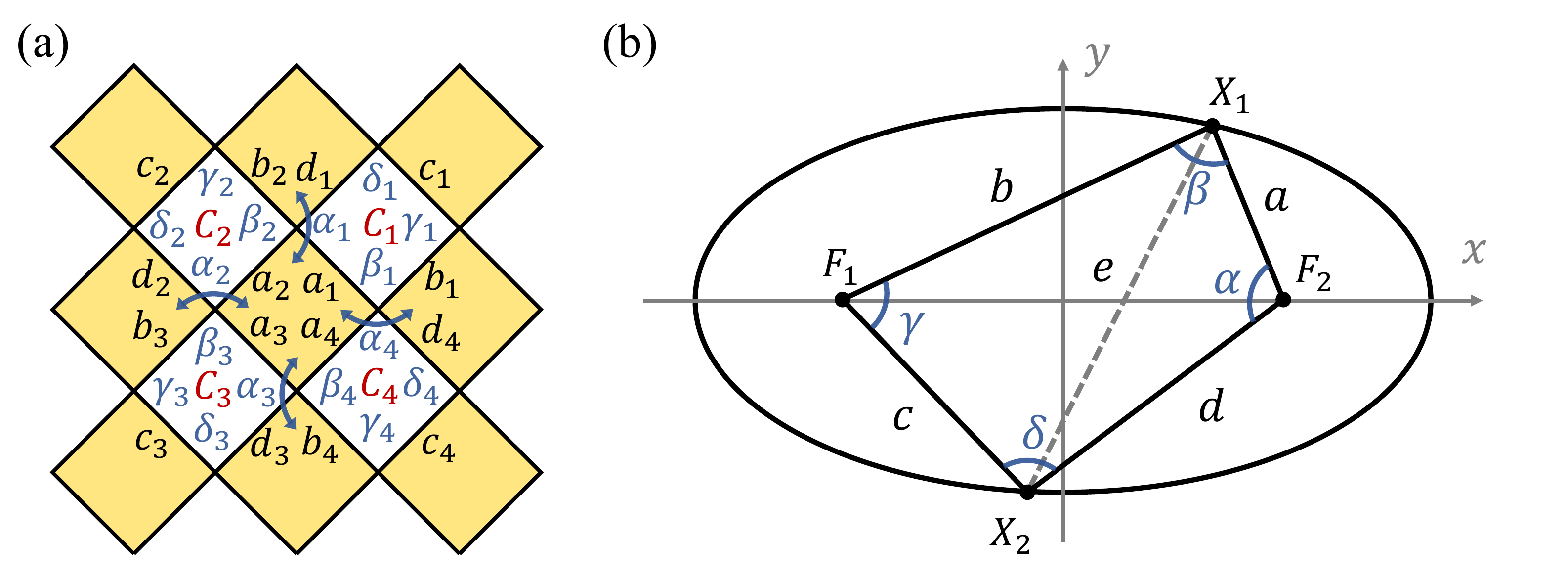}
	\caption{(a) Geometric notations of a $3\times3$ genus-0 PQK tessellation. The blue arrows indicate the expanding directions of opening angles. (b) A quadrilateral determined by two foci and two vertices on an ellipse.}
	\label{fig:lemma1}
\end{figure}

We begin the proof of Lemma~\ref{lemma1} by examining the basic $3\times3$ PQK tessellations as illustrated in Fig.~\ref{fig:lemma1}(a).
The side lengths of the head-to-tail connected quadrilateral voids $C_i$ are denoted by $a_i$, $b_i$, $c_i$, and $d_i$, and the opening angles are denoted by $\alpha_i$, $\beta_i$, $\gamma_i$, and $\delta_i\in[0,\pi]$ for $i=1,2,3,4$.
Since these voids are straight segments at the undeployed state (i.e., $\alpha_i=\gamma_i=0$, $\beta_i=\delta_i=\pi$), we have the constraints $a_i+b_i=c_i+d_i$.
Therefore, the quadrilaterals formed by the voids of cuts can be parametrized by an ellipse as shown in Fig.~\ref{fig:lemma1}(b).
Besides, the relationships $\alpha_i+\beta_{i+1}=\pi$ for opening angles of adjacent quadrilateral voids $C_i$ and $C_{i+1}$ hold upon deployment, where $i$ cycles from 1 to 4 (the same below).
We denote the functions $g_i$ as the relationship between cosines of opening angles $\cos\beta_i$ and $\cos\beta_{i+1}$:
\begin{equation}
	\cos\beta_i=g_i(\cos\beta_{i+1}) \triangleq g[\cos\beta_{i+1};a_i,b_i,c_i,d_i],
\end{equation}
where the function $g$ is defined by
\begin{equation}
	\begin{aligned}
		&\cos\beta=g(-\cos\alpha;a,b,c,d) \triangleq \\ &\cos\left[\arccos\left(\frac{a^2+e^2-d^2}{2ae}\right)+\arccos\left(\frac{b^2+e^2-c^2}{2be}\right)\right]
	\end{aligned}
\end{equation}
and
\begin{equation}
	e = \sqrt{a^2+d^2-2ad\cos\alpha}.
\end{equation}
The expression of $g$ is actually obtained from the relationship $\beta=\angle F_1X_1X_2+\angle F_2X_1X_2$ and $e=X_1X_2$ as illustrated in Fig.~\ref{fig:lemma1}(b).
The functions $g_i$ reflect on the dependence between the opening angles of adjacent voids of cuts.
Thus, a valid deployed configuration for $3\times3$ PQK tessellations requires the following loop condition:
\begin{equation}
	g_1 \circ g_2 \circ g_3 \circ g_4(\cos\beta_1) \equiv \cos\beta_1.
	\label{eq:loop}
\end{equation}
Here the operator $\circ$ represents the composition of two functions defined by $g_i\circ g_{i+1}(x)=g_i[g_{i+1}(x)]$.
This condition guarantees that the opening angle $\beta_1$ does not change through a loop of the quadrilateral voids by the relationship $\alpha_i=\pi-\beta_{i+1}$.

Obviously, if all the voids of cuts are parallelograms, we have $\cos\beta_i=\cos\beta_{i+1}$, so that $g_1 \circ g_2 \circ g_3 \circ g_4$ is an identity function and Eq.~(\ref{eq:loop}) is always satisfied, and then the tessellation is rigidly deployable.
To prove Lemma~\ref{lemma1}, we need to verify that Eq.~(\ref{eq:loop}) does not hold if there are nonparallelogram voids of cuts.
Because the expressions of $g_i$ are complicated to deal with for arbitrary quadrilaterals, we investigate the differential form of Eq.~(\ref{eq:loop}) instead:
\begin{equation}
	g'_1(\cos\beta_2) \cdot g'_2(\cos\beta_3) \cdot g'_3(\cos\beta_4) \cdot g'_4(\cos\beta_1) \equiv 1,
	\label{eq:dloop}
\end{equation}
where $g'_i$ can be calculated by
\begin{equation}
	g'_i(\cos\beta_{i+1}) =g'[\cos\beta_{i+1};a_i,b_i,c_i,d_i].
\end{equation}
In order to derive the expression of $g'=-{\dif\cos\beta}/{\dif\cos\alpha}$, we combine the following differential equations of quadrilaterals
\begin{equation}
	\begin{aligned}
		& ad\sin\alpha\dif\alpha = bc\sin\gamma\dif\gamma, \\
		& ab\sin\beta\dif\beta = cd\sin\delta\dif\delta, \\
		& \dif\alpha + \dif\beta + \dif\gamma + \dif\delta = 0,
	\end{aligned}
\end{equation}
and the identity
\begin{equation}
	ad\sin\alpha+bc\sin\gamma = ab\sin\beta+cd\sin\delta.
\end{equation}
Then the differential relationship between the adjacent opening angles $\alpha$ and $\beta$ is obtained as
\begin{equation}
	\frac{\dif\beta}{\dif\alpha} = - \frac{d\sin\delta}{b\sin\gamma}.
\end{equation}
Thus, we have
\begin{equation}
	g'(-\cos\alpha;a,b,c,d) =  \frac{d\sin\beta\sin\delta}{b\sin\alpha\sin\gamma}.
\end{equation}
Naturally, Eq.~(\ref{eq:dloop}) is a necessary condition of Eq.~(\ref{eq:loop}). Next we prove that $g'$ is monotonic when there are nonparallelogram voids.
By direct calculations, the second derivative of $g$ is
\begin{equation}
	\begin{aligned}
		g''(& -\cos\alpha;\; a,b,c,d) =
		k_g ( -a d \sin \alpha  \cot \gamma \\ & -a b \sin \beta  \cot \delta -b c  \sin \gamma \cot \alpha -c d  \sin \delta \cot \beta ),
	\end{aligned}
	\label{eq:ddf}
\end{equation}
where $k_g=d\sin\beta\sin\delta/(b^2c\sin^2\alpha\sin^2\gamma)$, and we have $k_g>0$ for $\alpha,\beta,\gamma,\delta\in(0,\pi)$.
Note that a quadrilateral formed by the voids of cuts can be determined by an ellipse as illustrated in Fig.~\ref{fig:lemma1}(b).
We denote the focal distance by $2f$, and length of the major axis by $2l$ with $0<f<l$.
Hence, Eq.~(\ref{eq:ddf}) can be written in terms of the coordinates of the foci $F_1=(-f,0,0)$, $F_1=(f,0,0)$, and two points $X_1=(l\cos\theta_1,\sqrt{l^2-f^2}\sin\theta_1,0)$, $X_2=(l\cos\theta_2,\sqrt{l^2-f^2}\sin\theta_2,0)$ on the ellipse:
\begin{equation}
	g''=k_g(f^2-l^2)\frac{1+\cos(\theta_1-\theta_2)}{\sin\theta_1\sin\theta_2}\frac{t_1f^2+t_2l^2}{t_3f^2+t_4l^2},
	\label{eq:ddfp}
\end{equation}
where the coefficients are defined by
\begin{equation}
	\begin{aligned}
		& t_1=- \sin ^2(\theta_1+\theta_2),\\
		& t_2= 2 \sin ^2\theta_1-\sin ^2(\theta_1-\theta_2)+2 \sin ^2\theta_2,\\
		& t_3=1+\cos(\theta_1+\theta_2),\\
		& t_4=-1-\cos(\theta_1-\theta_2),
	\end{aligned}
\end{equation}
and the parameters $\theta_1\in(0,\pi)$ and $\theta_2\in(-\pi,0)$ represent two points above and below the $x$ axis, respectively.
The sign of $g''$ can be determined by the following considerations.
First, for a quadrilateral that is not a parallelogram, we have $\theta_1-\theta_2\neq\pi$, so that $k_g(f^2-l^2){[1+\cos(\theta_1-\theta_2)]}/({\sin\theta_1\sin\theta_2})>0$.
Second, the voids of cuts are convex quadrilaterals; thus, the following constraints should be satisfied:
\begin{equation}
	\begin{aligned}
		& \bfe_3\cdot[(X_1-F_2)\times(X_2-F_2)]>0,\\
		& \bfe_3\cdot[(X_2-F_1)\times(X_1-F_1)]>0,
	\end{aligned}
	\label{eq:rangel}
\end{equation}
where $\bfe_3=(0,0,1)$.
Then we can obtain the range $l^2\in(f^2,-t_3/t_4f^2)$ from Eq.~(\ref{eq:rangel}), implying $t_3f^2+t_4l^2>0$.
Third, we have $t_2>0$ and $t_1f^2+t_2l^2> t_1f^2+t_2f^2=4\sin^2\theta_1\sin^2\theta_2>0$.
Altogether, we conclude that $g''>0$ for a nonparallelogram quadrilateral on an ellipse, and therefore, the function $g'$ increases monotonically.
For the $3\times3$ tessellation illustrated in Fig.~\ref{fig:lemma1}(a), on the one hand, with regard to the voids of cuts that are not parallelograms (e.g., $C_i$), the function $g'_i$ is always positive and increases monotonically; on the other hand, for the parallelogram voids of cuts (e.g., $C_j$), we have $g'_j\equiv1$.
As a result, if  any of the voids are not parallelograms, we will find that Eq.~(\ref{eq:dloop}) does not hold, which means that the loop condition Eq.~(\ref{eq:loop}) is violated.
Since a larger genus-0 PQK tessellation is composed of $3\times3$ parts, the rigid deployability thereof requires that all the voids of cuts are parallelograms.
Besides, if the voids of cuts are all parallelograms for a genus-0 PQK tessellation with arbitrary number of panels, we can verify that Eq.~(\ref{eq:loop}) holds for every loop of the voids, so that the tessellation is rigid deployable, Q.E.D.

\section{Proof of  Lemma~\ref{lemma2}}
\label{ap:lemma2}

\begin{figure}[!t]
	\centering
	\includegraphics[width=0.5\textwidth]{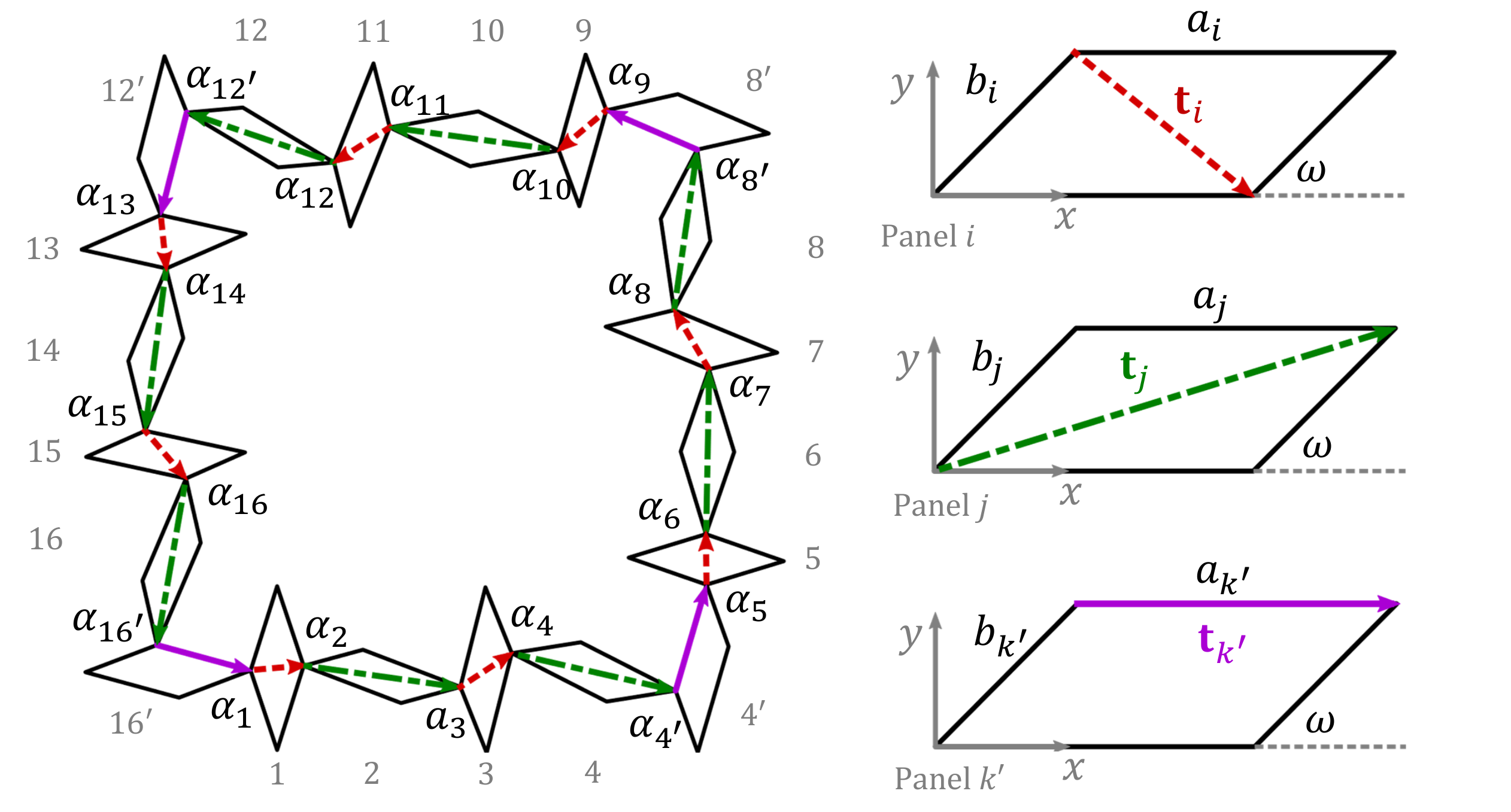}
	\caption{Ring of parallelogram voids. The minor diagonals (red dotted arrow), major diagonals (green dash-dotted arrow), and edges (purple solid arrow) of corresponding parallelograms are connected head to tail and form a closed ring of vectors.}
	\label{fig:lemma2}
\end{figure}

The geometrical notations of the ring of parallelograms are illustrated in Fig.~\ref{fig:lemma2}.
Our goal is to prove that this ring remains closed when the opening angles of these parallelograms vary simultaneously.
Without loss of generality, we assume that there are totally $m+4$ parallelograms ($m$ is an even number) including four parallelograms on the corner indexed by the even numbers $k'_1<k'_2<k'_3<k'_4=m'$.
Specifically, we have $k'_1=4'$, $k'_2=8'$, $k'_3=12'$, $k'_4=16'$, and $m=16$ for the example in Fig.~\ref{fig:lemma2}.
Since the ring is extracted from a given genus-1 PQK tessellation, the included angles between these parallelograms (i.e., $\alpha_i$ and $\alpha'_j$) are fixed.
It can be seen that the closed ring is essentially determined by the vectors of minor diagonals (red dotted arrow), major diagonals (green dash-dotted arrow), and edges (purple solid arrow) of the parallelograms.
The coordinates of these vectors can be expressed by
\begin{equation}
	\begin{aligned}
		&\bft_i^\omega = (a_i-b_i\cos\omega,-b_i\sin\omega)^{\rm T}, & & i=1,3,5,...,m-1, \\
		&\bft_j^\omega = (a_j+b_j\cos\omega,b_j\sin\omega)^{\rm T}, & & j=2,4,6,...,m, \\
		&\bft_{k}^\omega = (a_{k},0)^{\rm T}, & & k=k'_1,k'_2,k'_3,k'_4,
	\end{aligned}
	\label{eq:vector}
\end{equation}
where $a_i$, $b_i$, $a_j$, $b_j$, $a_k$, and $b_k$ are the side lengths of the parallelograms, as illustrated in Fig.~\ref{fig:lemma2} right.
Similar to the parallelograms, these vectors are also connected head-to-tail with rotations by angles $\alpha_i$ and $\alpha'_i$ determined by the given kirigami pattern.
The condition that these vectors form a closed loop is as follows:
\begin{equation}
	\begin{aligned}
		\bfR^{\alpha_1}\bft_1^\omega&+\bfR^{\alpha_1}\bfR^{-\alpha_2}\bft_2^\omega+\bfR^{\alpha_1}\bfR^{-\alpha_2}\bfR^{\alpha_3}\bft_3^\omega+\cdots\\
		&+\bfR^{\alpha_1}\bfR^{-\alpha_2}\bfR^{\alpha_3}\cdots\bfR^{-\alpha_{k_1}}\bft_{k_1}^\omega\\
		&+\bfR^{\alpha_1}\bfR^{-\alpha_2}\bfR^{\alpha_3}\cdots\bfR^{-\alpha_{k_1}}\bfR^{\alpha_{k'_1}}\bft_{k'_1}^\omega+\cdots\\
		&+\bfR^{\alpha_1}\bfR^{-\alpha_2}\bfR^{\alpha_3}\cdots\bfR^{-\alpha_m}\bft_m^\omega\\
		&+\bfR^{\alpha_1}\bfR^{-\alpha_2}\bfR^{\alpha_3}\cdots\bfR^{-\alpha_m}\bfR^{\alpha_{m'}}\bft_{m'}^\omega={\bf 0}.
	\end{aligned}
	\label{eq:ring-0}
\end{equation}
For simplicity, we define the following rotation angles
\begin{equation}
	\begin{aligned}
		&\beta_1=\alpha_1,\\
		&\beta_2=\alpha_1-\alpha_2,\\
		&\beta_3=\alpha_1-\alpha_2+\alpha_3,\\
		&~~~~\cdots\\
		&\beta_{k_1}=\alpha_1-\alpha_2+\alpha_3+\cdots-\alpha_{k_1},\\
		&\beta_{k'_1}=\alpha_1-\alpha_2+\alpha_3+\cdots-\alpha_{k_1}+\alpha_{k'_1},\\
		&~~~~~\cdots\\
		&\beta_m=\alpha_1-\alpha_2+\alpha_3+\cdots-\alpha_{k_1}+\alpha_{k'_1}+\cdots-\alpha_m,\\
		&\beta_{m'}=\alpha_1-\alpha_2+\alpha_3+\cdots-\alpha_{k_1}+\alpha_{k'_1}+\cdots-\alpha_m+\alpha_{m'}.\\
	\end{aligned}
	\label{eq:ring-1-beta}
\end{equation}
Then Eq.~(\ref{eq:ring-0}) can be rewritten as
\begin{equation}
	\begin{aligned}
		\bfR^{\beta_1}\bft_1^\omega&+\bfR^{\beta_2}\bft_2^\omega+\bfR^{\beta_3}\bft_3^\omega+\cdots\\
		&+\bfR^{\beta_{k_1}}\bft_{k_1}^\omega
		+\bfR^{\beta_{k'_1}}\bft_{k'_1}^\omega+\cdots\\
		&+\bfR^{\beta_m}\bft_m^\omega
		+\bfR^{\beta_{m'}}\bft_{m'}^\omega={\bf 0}.
	\end{aligned}
	\label{eq:ring-1}
\end{equation}
Furthermore, Eq.~(\ref{eq:ring-1}) can be arranged in terms of $\cos\omega$ and $\sin\omega$ as follows:
\begin{equation}
	\begin{aligned}
		&u_1+u_2\cos\omega+u_3\sin\omega=0,\\
		&\hat{u}_1-u_3\cos\omega+u_2\sin\omega=0,
	\end{aligned}
	\label{eq:ring-2}
\end{equation}
where the parameters $u_1$, $\hat{u}_1$, $u_2$, and $u_3$ are defined by
\begin{equation}
	\begin{aligned}
		&u_1=\sum\nolimits_{i=1}^m (a_i\cos\beta_i)+\sum\nolimits_{i=1}^4 (a_{k'_i}\cos\beta_{k'_i}),\\
		&\hat{u}_1=\sum\nolimits_{i=1}^m (a_i\sin\beta_i)+\sum\nolimits_{i=1}^4 (a_{k'_i}\sin\beta_{k'_i}),\\
		&u_2=\sum\nolimits_{i=1}^m [(-1)^i b_i\cos\beta_i],\\
		&u_3=\sum\nolimits_{i=1}^m [(-1)^{i-1} b_i\sin\beta_i].
	\end{aligned}
	\label{eq:ring-2-u-1}
\end{equation}
These parameters are independent of the reference opening angle $\omega$.
Considering that the closed loop of vectors does exist for the undeployed configuration ($\omega=0$) of any kirigami pattern, we obtain
\begin{equation}
	\begin{aligned}
		&u_1+u_2=0,\\
		&\hat{u}_1-u_3=0.
	\end{aligned}
	\label{eq:ring-2-u-2}
\end{equation}
Thus, Eq.~(\ref{eq:ring-2}) is equivalent to the matrix form
\begin{equation}
	\bfM^\omega\bfu=0,
	\label{eq:ring-3}
\end{equation}
where the coefficient matrix $\bfM^\omega$ and vector $\bfu$ are
\begin{equation}	
	\bfM^\omega=
	\left[
	\begin{array}{cc}
		1-\cos\omega &   \sin\omega\\
		-\sin\omega & 1-\cos\omega
	\end{array}
	\right],
	\bfu=
	\left[
	\begin{array}{c}
		u_1\\
		\hat{u}_1
	\end{array}
	\right].
	\label{eq:ring-3-2}
\end{equation}
The determinant of $\bfM^\omega$ is
\begin{equation}
	{\rm det}(\bfM^\omega)=2(1-\cos\omega).
	\label{eq:ring-3-det}
\end{equation}

Now we go back to Lemma~\ref{lemma2}.
On the one hand, for a given genus-1 PQK tessellation, if there exists a deployed configuration with parallelogram voids of cuts, we have a specific ${\tilde\omega}\in(0,\pi]$, such that $\bfM^{{\tilde\omega}}\bfu=0$.
Since ${\rm det}(\bfM^{{\tilde\omega}})>0$ for this situation, we have $\bfu=0$.
Therefore, for any $\omega\in[0,\pi]$, Eq.~(\ref{eq:ring-3}) always holds, which means that this genus-1 PQK tessellation is rigidly deployable.
On the other hand, it is trivial that an RDPQK genus-1 tessellation has valid deployed configurations, and the voids of cuts must be parallelograms to guarantee the rigid deployability as stated in Sec.~\ref{sect:genus1}, Q.E.D.

\begin{widetext}
\section{{Matrix of topology}}
The {matrices of topology} of the tessellations illustrated in Figs.~\ref{fig:intro}(b) and \ref{fig:genus4}(b) are given below:
	\begin{equation}
		\bfT^1_{10\times10}=\left[
		\begin{array}{rrrrrrrrrrr}
			-1 &     -2 &     -1 &     -2 &     -1 &     -2 &     -1 &     -2 &     -1 &     -2 &     -1 \\
			-2 &      1 &      2 &      1 &      2 &      1 &      2 &      1 &      2 &      1 &     -2 \\
			-1 &      2 &      1 &      2 &      1 &      2 &      1 &      2 &      1 &      2 &     -1 \\
			-2 &      1 &      2 &     -1 &     -2 &     -1 &     -2 &     -1 &      2 &      1 &     -2 \\
			-1 &      2 &      1 &     -2 &      0 &      0 &      0 &     -2 &      1 &      2 &     -1 \\
			-2 &      1 &      2 &     -1 &      0 &      0 &      0 &     -1 &      2 &      1 &     -2 \\
			-1 &      2 &      1 &     -2 &      0 &      0 &      0 &     -2 &      1 &      2 &     -1 \\
			-2 &      1 &      2 &     -1 &     -2 &     -1 &     -2 &     -1 &      2 &      1 &     -2 \\
			-1 &      2 &      1 &      2 &      1 &      2 &      1 &      2 &      1 &      2 &     -1 \\
			-2 &      1 &      2 &      1 &      2 &      1 &      2 &      1 &      2 &      1 &     -2 \\
			-1 &     -2 &     -1 &     -2 &     -1 &     -2 &     -1 &     -2 &     -1 &     -2 &     -1
		\end{array}
		\right],
		\label{eq:tm10}
	\end{equation}
	
	\begin{equation}
		\bfT^4_{17\times17}=\left[
		\begin{array}{rrrrrrrrrrrrrrrrrr}
			-2 &     -1 &     -2 &     -1 &     -2 &     -1 &     -2 &     -1 &     -2 &     -1 &     -2 &     -1 &     -2 &     -1 &     -2 &     -1 &     -2 &     -1 \\
			-1 &      2 &      1 &      2 &      1 &      2 &      1 &      2 &      1 &      2 &      1 &      2 &      1 &      2 &      1 &      2 &      1 &     -2 \\
			-2 &      1 &      2 &      1 &      2 &      1 &      2 &      1 &      2 &      1 &      2 &      1 &      2 &      1 &      2 &      1 &      2 &     -1 \\
			-1 &      2 &      1 &     -2 &     -1 &     -2 &     -1 &     -2 &      1 &      2 &     -1 &     -2 &     -1 &     -2 &     -1 &      2 &      1 &     -2 \\
			-2 &      1 &      2 &     -1 &      0 &      0 &      0 &     -1 &      2 &      1 &     -2 &      0 &      0 &      0 &     -2 &      1 &      2 &     -1 \\
			-1 &      2 &      1 &     -2 &      0 &      0 &      0 &     -2 &      1 &      2 &     -1 &      0 &      0 &      0 &     -1 &      2 &      1 &     -2 \\
			-2 &      1 &      2 &     -1 &      0 &      0 &      0 &     -1 &      2 &      1 &     -2 &      0 &      0 &      0 &     -2 &      1 &      2 &     -1 \\
			-1 &      2 &      1 &     -2 &     -1 &     -2 &     -1 &     -2 &      1 &      2 &     -1 &     -2 &     -1 &     -2 &     -1 &      2 &      1 &     -2 \\
			-2 &      1 &      2 &      1 &      2 &      1 &      2 &      1 &      2 &      1 &      2 &      1 &      2 &      1 &      2 &      1 &      2 &     -1 \\
			-1 &      2 &      1 &      2 &      1 &      2 &      1 &      2 &      1 &      2 &      1 &      2 &      1 &      2 &      1 &      2 &      1 &     -2 \\
			-2 &      1 &      2 &     -1 &     -2 &     -1 &     -2 &     -1 &      2 &      1 &     -2 &     -1 &     -2 &     -1 &     -2 &      1 &      2 &     -1 \\
			-1 &      2 &      1 &     -2 &      0 &      0 &      0 &     -2 &      1 &      2 &     -1 &      0 &      0 &      0 &     -1 &      2 &      1 &     -2 \\
			-2 &      1 &      2 &     -1 &      0 &      0 &      0 &     -1 &      2 &      1 &     -2 &      0 &      0 &      0 &     -2 &      1 &      2 &     -1 \\
			-1 &      2 &      1 &     -2 &      0 &      0 &      0 &     -2 &      1 &      2 &     -1 &      0 &      0 &      0 &     -1 &      2 &      1 &     -2 \\
			-2 &      1 &      2 &     -1 &     -2 &     -1 &     -2 &     -1 &      2 &      1 &     -2 &     -1 &     -2 &     -1 &     -2 &      1 &      2 &     -1 \\
			-1 &      2 &      1 &      2 &      1 &      2 &      1 &      2 &      1 &      2 &      1 &      2 &      1 &      2 &      1 &      2 &      1 &     -2 \\
			-2 &      1 &      2 &      1 &      2 &      1 &      2 &      1 &      2 &      1 &      2 &      1 &      2 &      1 &      2 &      1 &      2 &     -1 \\
			-1 &     -2 &     -1 &     -2 &     -1 &     -2 &     -1 &     -2 &     -1 &     -2 &     -1 &     -2 &     -1 &     -2 &     -1 &     -2 &     -1 &     -2
		\end{array}
		\right].
		\label{eq:tm17}
	\end{equation}
\end{widetext}

%

\end{document}